\documentclass[journal]{IEEEtran}
\usepackage{amsmath,amsfonts}
\usepackage{algorithmic}
\usepackage[ruled,linesnumbered]{algorithm2e}
\usepackage{array}
\usepackage[caption=false,font=normalsize,labelfont=sf,textfont=sf]{subfig}
\usepackage{textcomp}
\usepackage{stfloats}
\usepackage{url}
\usepackage{balance}
\usepackage{verbatim}
\usepackage{graphicx}
\graphicspath{{figures/}}
\usepackage{cite}
\usepackage[colorlinks]{hyperref}
\hypersetup{citecolor=blue}
\usepackage{threeparttable}
\usepackage{multirow}
\usepackage{booktabs}
\usepackage{amssymb}
\usepackage{bm}
\usepackage{color}
\usepackage{mathtools}

\newcommand{\trans}{{\mathrm{T}}}
\newcommand{\herm}{{\mathrm{H}}}

\newcommand{\C}{\mathbb{C}}
\newcommand{\R}{\mathbb{R}}

\DeclareMathOperator{\Real}{Re}

\begin{document}

\title{Flexible Beamforming via Fluid Antenna Array: From Discrete to Continuous Array}

\author{Jingyuan Xu, Yuanhui Wu, Zhentian Zhang,~\IEEEmembership{Member,~IEEE}, Zaichen Zhang,~\IEEEmembership{Senior Member,~IEEE}, \\Tuo Wu, Hao Jiang,~\IEEEmembership{Senior Member,~IEEE}, Jian Dang,~\IEEEmembership{Senior Member,~IEEE},\\Christos Masouros,~\IEEEmembership{Fellow,~IEEE}, Kin-Fai Tong,~\IEEEmembership{Fellow,~IEEE}
\vspace{-8mm}
   	\thanks{Jingyuan Xu and Zaichen Zhang are with the National Mobile Communications Research Laboratory, Frontiers Science Center for Mobile Information Communication and Security, Southeast University, Nanjing, 210096, China. Zaichen Zhang is also with the Purple Mountain Laboratories, Nanjing 211111, China. (e-mails: \{jingyuanxu, zczhang\}@seu.edu.cn).}
   	\thanks{Yuanhui Wu is with the College of Artificial Intelligence, Nanjing University of Information Science and Technology, 210044, China. (e-mails: 202412621447@nuist.edu.cn).}
   \thanks{Zhentian Zhang is with the Department of Electrical and Electronic Engineering, The Hong Kong Polytechnic University, Hong Kong, SAR (e-mail: zhentianzhangzzt@gmail.com).}
	\thanks{H. Jiang is with the School of Cyber Science and Engineering, Southeast University, Nanjing 210096, P. R. China (e-mail: jiang.hao@seu.edu.cn).}
	\thanks{Jian Dang is with the National Mobile Communications Research Laboratory, Frontiers Science Center for Mobile Information Communication and Security, Southeast University, Nanjing 211189, China, also with the Key Laboratory of Intelligent Support Technology for Complex Environments, Ministry of Education, Nanjing University of Information Science and Technology, Nanjing 210044, China, and also with Purple Mountain Laboratories, Nanjing 211111, China (email: dangjian@seu.edu.cn).} 
	\thanks{Christos Masouros is with the Department of Electronic and Electrical Engineering, University College London, Torrington Place, WC1E 7JE, United Kingdom (e-mail: c.masouros@ucl.ac.uk).}
	\thanks{Tuo Wu is with the School of Electronic and Information Engineering, South China University of Technology, Guangzhou 510640, China (E-mail: wutuo@scut.edu.cn).}
	\thanks{Kin-Fai Tong is with the School of Science and Technology, Hong Kong Metropolitan University, Hong Kong, China (e-mail: ktong@hkmu.edu.hk).}
	\thanks{Corresponding Author: Zhentian Zhang}
}

\maketitle
\begin{abstract}
Fluid antenna arrays (FAAs) reconfigure their effective aperture by activating radiation ports from a dense set of candidate positions within a finite physical region, thereby offering geometry diversity beyond conventional fixed arrays. This paper studies arbitrary-direction far-field beam synthesis with a planar FAA. Since a fixed array adapts its beamforming weights on a predetermined spatial sampling structure, its ability to match desired beam patterns is limited when both the aperture and the number of active elements are constrained. To exploit the geometry diversity of the FAA, we propose a coarse-to-fine scheme that couples discrete port selection with continuous joint refinement. In the coarse stage, a fast Fourier transform (FFT) based phase retrieval step converts the amplitude-only beam specification into an aperture-compatible complex target, after which an FFT-tailored greedy method selects the active ports and least squares initializes their complex weights. In the fine stage, a projected adaptive moment estimation procedure jointly refines the continuous port positions and beamforming weights while enforcing the minimum interelement spacing. For the evaluated two-beam configuration, the proposed scheme increases the mainlobe energy fraction from 35.73\% to 48.92\% and reduces the peak sidelobe level from -2.30~dB to -3.83~dB relative to a fixed array with the same aperture and active-port count.
\end{abstract}


\section{Introduction}\label{sec:intro}
\subsection{Background and Related Work}
\begin{figure}[!t]
	\centering
	\includegraphics[width=\columnwidth]{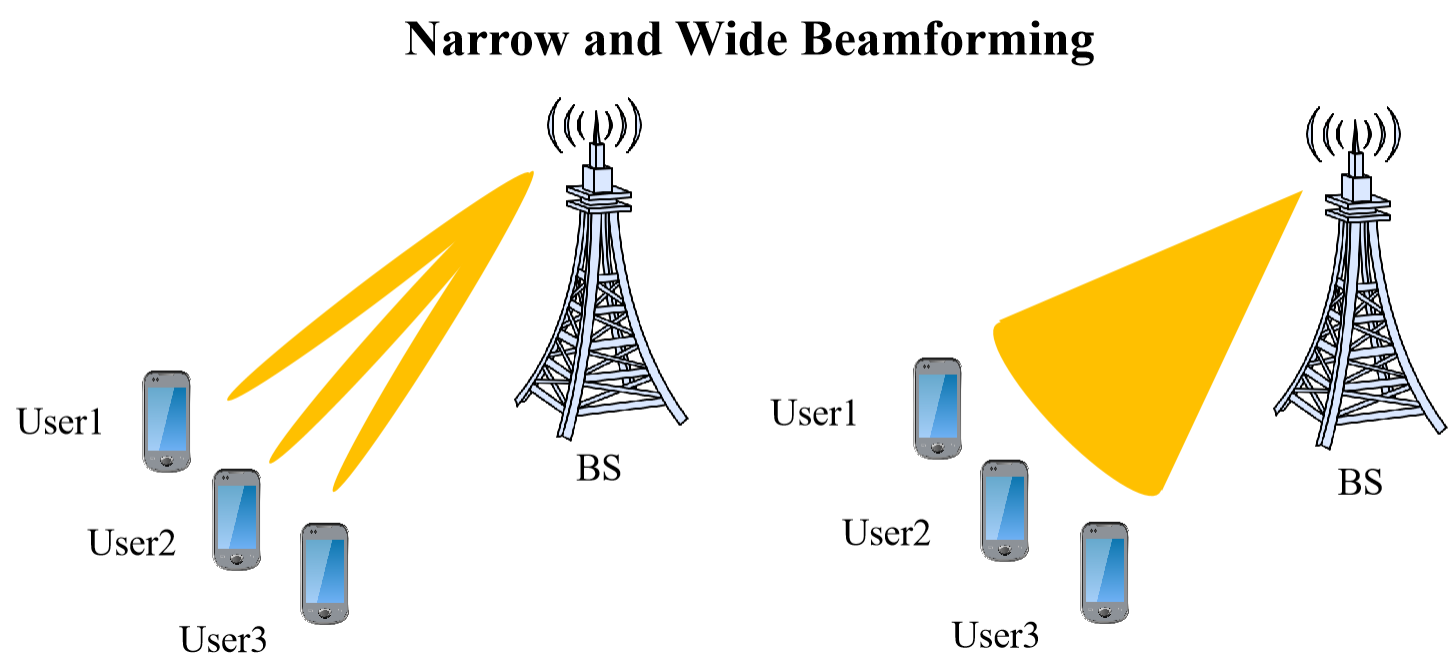}
\caption{Illustrations of narrow- and wide-beam coverage in wireless communications.}
	\label{fig:Wide}
\end{figure}

\IEEEPARstart{B}{eamforming} controls the angular distribution of an array's field through the coherent superposition of waves radiated by its elements. Its design specifies the beam directions, widths, coverage regions, and sidelobe levels. Narrow beams concentrate the array response around selected directions~\cite{NB1}, while broad beams provide coverage over angular sectors~\cite{NB2}. Fig.~\ref{fig:Wide} illustrates these two coverage patterns in a communication setting. Applications such as directional transmission and sector-wide broadcasting motivate an array architecture that synthesizes beams with different pointing directions and widths within a common framework.

Fluid antenna systems (FASs) \cite{fas_wong2020,fas_wong2021,fbl_fas} offer precisely such flexibility at the antenna level: the active radiating elements are no longer anchored at fixed locations but can be relocated or switched among candidate positions inside a finite aperture~\cite{fas_tutorial}. The word ``fluid'' refers to this softwarized adaptability rather than to liquid antenna materials, so reconfigurable pixel antennas and selectable-port arrays also fall under the FAS umbrella once the effective radiation position or aperture response becomes electronically controllable~\cite{fas_pixel}. Most early FAS works concentrated on the spatial variation of the channel across the aperture, often referred to as {\em fading diversity} \cite{zzt2,zzt1,wt1,wt2,HJ1,HJ2,zzt3,zzt4,wt3,wt4,zzt5,zzt6,HJ3,zzt7,wt5,wt6,wt7,wt8}: moving the active port to a favorable position strengthens the received signal, enhances reliability, and suppresses interference.

Fluid antenna arrays (FAAs) \cite{faa_spacing_linear,faa_spacing_planar,gd3} push this idea further by activating multiple ports simultaneously. The chosen port layout then dictates not only the channel samples but also the effective aperture, the spatial sampling density, the interelement spacings, and hence the entire array manifold, all of which shape the far-field radiation pattern. We term this controllable spatial capability {\em geometry diversity}~\cite{faa_spacing_linear,faa_spacing_planar,omp_conference,gd2,gd3,gd4,gd5}. Whereas fading diversity originates from the random propagation environment, geometry diversity stems from the reconfigurable structure of the array itself and is therefore deterministic: it can be analyzed, optimized, and matched to the radiation objective in advance. With a fixed active-port layout, an array adapts its radiation pattern through the complex beamforming weights on a prescribed spatial sampling structure. Jointly adapting the active-port layout and the complex weights provides additional design freedom for matching the desired beam pattern within a fixed aperture and active-port budget. {\em Designing the port activation of an FAA is thus a geometry-aware array synthesis problem rather than merely a channel selection problem.}

Classical antenna selection (AS) connects a subset of fixed antenna elements to a limited number of radio frequency (RF) chains, trading hardware complexity against communication performance~\cite{as_molisch2004}. Hybrid analog/digital architectures distribute beamforming between baseband processing and an RF network implemented with phase shifters or switches~\cite{hybrid_mendez2016}. AS has been combined with hybrid precoding through joint optimization of the transmit-antenna subset and the analog and digital precoders for achievable-rate maximization~\cite{as_hybrid_liu2022}. FAA port selection shares the subset-selection structure of classical AS. This work synthesizes a prescribed far-field amplitude pattern within a finite planar aperture by combining phase retrieval, discrete port selection, and continuous refinement of port positions and complex weights. The active-port count is fixed, and the refined layout satisfies the minimum interelement spacing.

Selecting a small number of active ports from a dense candidate grid is inherently challenging because the number of possible configurations grows combinatorially with the grid size. Sparse approximation provides a tractable formulation of this problem, and greedy algorithms offer an attractive balance between computational efficiency and beam-synthesis accuracy~\cite{cs_candes,cs_donoho,omp_tropp}. Their performance, however, is sensitive to the correlation among candidate port responses and to the construction of the complex target field. Locally selected ports may be suboptimal, while an amplitude-only beam specification leaves the target phase undetermined. Although the feasibility of greedy port selection for planar FAA beamforming has been demonstrated in~\cite{omp_conference}, its potential for efficient and high-quality beam synthesis has yet to be fully exploited.

A complementary line of research treats element positions as continuous variables and jointly optimizes the coordinates and beamforming weights over a continuous region~\cite{ma_beamforming}, typically through gradient descent or alternating optimization. Without a structured discrete prior, however, such joint optimization is highly nonconvex and prone to highly suboptimal local minima. These observations motivate a coarse-to-fine design: discrete port selection first supplies a structured initialization, upon which continuous joint refinement further improves both the port positions and the weights.

\subsection{Challenges and Motivations}
Combining sparse port selection with continuous port refinement for an FAA requires resolving two issues: an amplitude-only beam specification does not determine a unique complex target for sparse synthesis, and continuous position refinement must reduce grid dependence while maintaining the minimum interelement spacing. The principal challenges are summarized below.

\begin{enumerate}
	\item \textbf{Phase Non-Uniqueness.} An amplitude or power specification does not uniquely determine the complex target field. Prescribing its phase in advance reduces the available design freedom and may compromise the sparsity achievable under the same power mask~\cite{morabito_phase_nonuniqueness}. Recovering a compatible phase from magnitude-only constraints is itself a nonconvex problem whose solution can depend on initialization~\cite{echeveste_phase_retrieval}. This motivates performing phase retrieval before sparse port selection to construct an aperture-adapted complex target.
	\item \textbf{Spacing-Constrained Off-Grid Refinement.} Grid-based sparse array synthesis faces a tradeoff between resolution and complexity: accurate nonuniform positions require extremely dense candidate grids, which incur high computational cost and may produce impractically small interelement spacings. Spacing-constrained off-grid refinement has been introduced to reduce the reliance on dense grids~\cite{yang_offgrid_cs}. However, minimum-distance constraints can partition the feasible set of port positions into disconnected subsets, causing sequential position updates to remain confined to the subset containing the initialization~\cite{jin_distance_constraints}. Therefore, all port coordinates should be jointly optimized, with the minimum interelement spacing explicitly restored after each update.
\end{enumerate}

\subsection{Contributions}
To address these challenges, this paper proposes a coarse-to-fine method that combines phase-retrieval-assisted discrete port selection with spacing-constrained continuous port refinement for arbitrary-direction beamforming with an FAA. The main contributions are summarized as follows.

\begin{enumerate}
	\item \textbf{Two-stage optimization framework.} We propose a structured pipeline in which an FFT-tailored orthogonal matching pursuit (OMP) first selects \(S\) ports from a discrete candidate grid, and least squares then initializes their complex beamforming weights. The resulting placement and weights provide a structured initial point for continuous joint optimization over the port coordinates and beamforming weights. This design connects combinatorial port selection with gradient-based continuous refinement and reduces the reliance on an extremely dense candidate grid.
	
	\item \textbf{Phase-retrieval-assisted target construction.} An FFT-based Gerchberg--Saxton phase retrieval step~\cite{gs1972} is integrated before port selection to convert the amplitude-only desired beam into a complex target with an aperture-adapted phase. This step improves the compatibility of the target supplied to OMP with the finite physical aperture.
	
\item \textbf{Adaptive moment estimation (Adam)-based continuous refinement.} All continuous port coordinates together with the real and imaginary parts of the beamforming weights are jointly refined using automatic differentiation and Adam~\cite{adam_kingma}. A common loss increases the mainlobe concentration while balancing the concentration across the desired beams. A minimum-spacing repair follows every position update, and the feasible iterate with the highest mainlobe concentration is retained.
\end{enumerate}

\textit{Notation:} Bold lowercase and uppercase letters denote vectors and matrices, respectively. The $(m,n)$-th entry and the $n$-th column of a matrix $\mathbf A$ are denoted by $[\mathbf A]_{m,n}$ and $[\mathbf A]_{:,n}$, respectively. For an index set \(\mathcal I\), \([\mathbf A]_{:,\mathcal I}\) denotes the submatrix formed by the columns of \(\mathbf A\) indexed by \(\mathcal I\). The superscripts $(\cdot)^{\trans}$ and $(\cdot)^{\herm}$ denote transpose and Hermitian transpose. During the iterations, $\mathbf A^{(t)}$ denotes the value of a variable at iteration $t$. The sets of real and nonnegative real numbers are denoted by \(\mathbb R\) and \(\mathbb R_+\). The pseudo-inverse is denoted by $(\cdot)^{\dagger}$. The Euclidean norm is $\|\cdot\|_2$, and $\|\cdot\|_0$ counts the number of nonzero entries. The Fourier transform and inverse Fourier transform are denoted by $\mathcal{F}\{\cdot\}$ and $\mathcal{F}^{-1}\{\cdot\}$.

\section{System Model}\label{sec:model}
\begin{figure*}[!t]
	\centering
	\includegraphics[width=\textwidth]{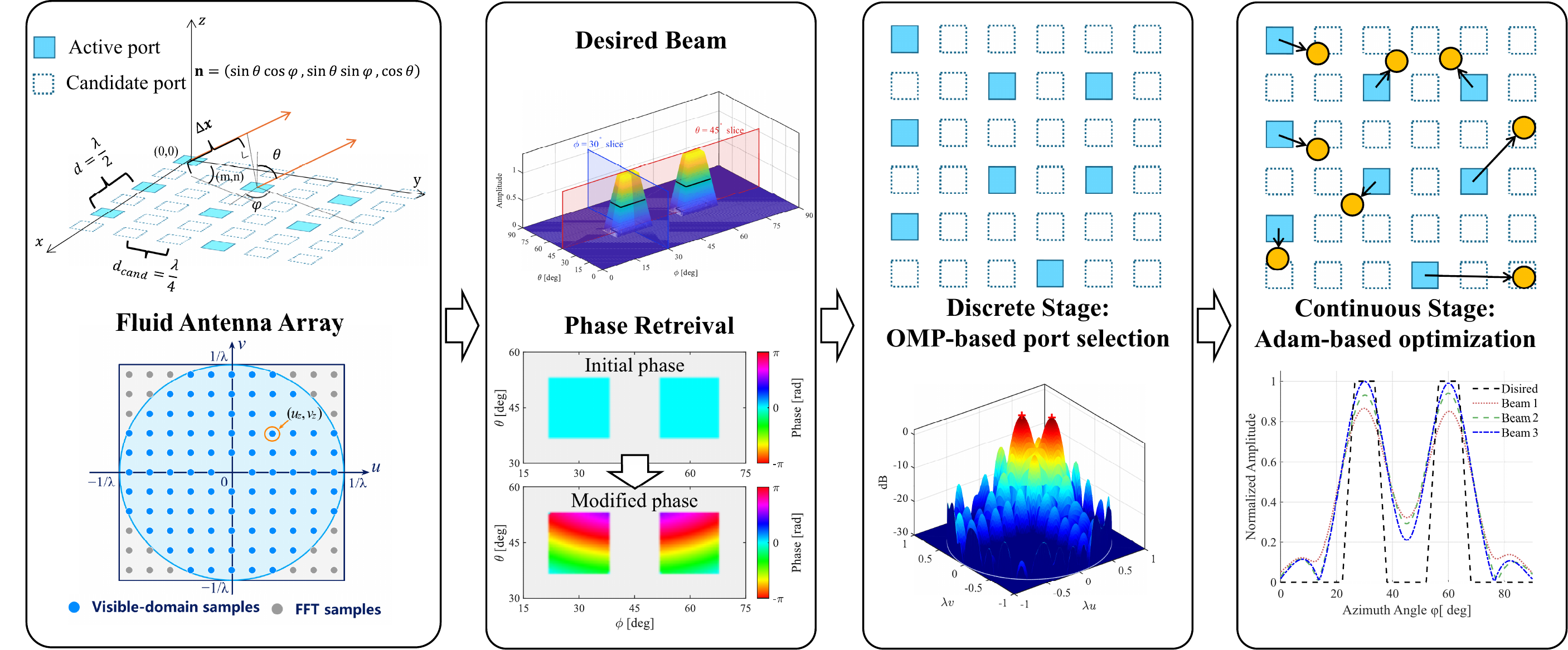}
	\caption{System model and overview of the proposed two-stage
		beamforming framework. A planar FAA radiates toward arbitrary
		far-field directions using a subset of candidate ports. The sampling
		diagram shows the uniform \((u,v)\) grid and the visible disk. FFT-based
		phase retrieval first converts the desired amplitude pattern into a
		complex target. OMP then selects active ports in the discrete stage,
		and projected Adam subsequently refines their continuous positions
		and beamforming weights.}
	\label{fig:framework-overview}
\end{figure*}

\subsection{Far-Field Beam Model for the FAA}
We consider narrowband far-field beam synthesis with a planar FAA.
The active-port coordinates and complex beamforming weights determine
the array response over the prescribed angular region. Under the
far-field approximation, the relative propagation delays across the
aperture are represented by direction-dependent phase shifts.

Consider a rectangular physical aperture of dimensions
\(W_x\times W_y\), aligned with the \(x\)- and \(y\)-axes. A total of
\(S\) ports are activated within this aperture. For a far-field
direction \((\theta,\varphi)\), we refer to
\(\theta\in[0,\pi/2]\) as the elevation angle for convenience. Here,
it is defined as the angle between the beam direction and the array
normal (the broadside direction). The azimuth angle
\(\varphi\in[0,2\pi)\) is measured from the \(x\)-axis in the array
plane. We map these angular coordinates to spatial-frequency variables
\begin{equation}
u=\frac{\sin\theta\cos\varphi}{\lambda},\quad
v=\frac{\sin\theta\sin\varphi}{\lambda},
\label{eq:uv_mapping}
\end{equation}
where \(\lambda\) denotes the carrier wavelength. The steering response
of a port at \((x,y)\) is \(e^{-j2\pi(xu+yv)}\). The visible hemisphere
maps to the disk
\(\mathcal V=\{(u,v):u^2+v^2\le 1/\lambda^2\}\).
The angular coordinates \((\theta,\varphi)\) describe physical directions,
and the \((u,v)\)-domain provides the sampling grid for beam synthesis
and performance evaluation.

The axis-aligned square \([-1/\lambda,1/\lambda]^2\) enclosing
\(\mathcal V\) is sampled on a uniform \(P\times Q\) grid in \((u,v)\). Let
\(\mathcal I_{\mathcal V}\) denote the indices that fall inside the
visible disk and let \(Z=|\mathcal I_{\mathcal V}|\). After retaining
these samples, the visible direction pairs are denoted by
\(\{(u_z,v_z)\}_{z=1}^{Z}\), and the corresponding desired complex beam
samples are collected in
\begin{equation}
\mathbf g=[g_1,\ldots,g_Z]^\trans\in\C^{Z\times1}.
\label{eq:desired_beam}
\end{equation}
Both the discrete and continuous formulations that follow share the
same aperture \(W_x\times W_y\), active-port count \(S\),
spatial-frequency grid \(\{(u_z,v_z)\}_{z=1}^{Z}\), and desired beam
\(\mathbf g\). The leftmost block of
Fig.~\ref{fig:framework-overview} illustrates the considered FAA
geometry, including the candidate ports, active ports, and far-field
direction.
The sampling diagram below the array shows a uniform FFT grid, with blue
points inside the visible disk and gray points outside it.

\subsection{Discrete Port-Selection Formulation}
In the discrete formulation, the aperture is covered by a dense
\(M_c\times N_c\) rectangular candidate grid. The spacing between
adjacent candidate ports in this grid is denoted by \(d_{\mathrm{cand}}\).
The grid coordinates are
\begin{equation}
x_i=(i-1)d_{\mathrm{cand}},\quad
y_j=(j-1)d_{\mathrm{cand}},
\label{eq:discrete_grid}
\end{equation}
for \(i=1,\ldots,M_c\) and \(j=1,\ldots,N_c\), such that
\(W_x=(M_c-1)d_{\mathrm{cand}}\) and
\(W_y=(N_c-1)d_{\mathrm{cand}}\). There are
\(N_p=M_cN_c\) candidate ports. After vectorizing
the grid in column-major order, the position of the \(l\)-th candidate
port is denoted by \(\mathbf r_l=[x_l,y_l]^\trans\) for
\(l=1,\ldots,N_p\).

The steering dictionary
\(\mathbf D\in\C^{Z\times N_p}\) contains the
far-field response of every candidate port, whose \((z,l)\)-th entry is
\begin{equation}
[\mathbf D]_{z,l}
=\exp[-j2\pi(x_lu_z+y_lv_z)].
\label{eq:discrete_dictionary}
\end{equation}
Let
\(\mathbf w\in\C^{N_p\times1}\) collect the
complex beamforming weights over the complete candidate grid. The
synthesized beam is
\begin{equation}
\mathbf y=\mathbf D\mathbf w.
\label{eq:discrete_beam}
\end{equation}
The nonzero entries of \(\mathbf w\) identify the active
ports and determine their complex beamforming weights. Discrete port selection and
beam synthesis can therefore be formulated as
\begin{equation}
\begin{gathered}
\min_{\mathbf w}~
    \|\mathbf g-\mathbf D\mathbf w\|_2^2\\
\mathrm{s.t.}~
    \|\mathbf w\|_0=S,\\
\|\mathbf r_l-\mathbf r_m\|_2\ge d_{\min},~
l,m\in\mathcal W,~l\ne m.
\end{gathered}
\label{eq:discrete_problem}
\end{equation}
where \(d_{\min}\) is the prescribed minimum inter-port distance and
\(\mathcal{W}=\{l:w_l\neq 0\}\) is the set of active ports. This is a
combinatorial sparse approximation problem, for which exhaustive search
is infeasible. In the reported setup, the candidate-grid spacing
satisfies \(d_{\mathrm{cand}}\ge d_{\min}\), so any subset of distinct grid points is
feasible and standard OMP can provide an initial placement. The
continuous refinement stage then maintains the aperture constraint and
corrects spacing violations introduced by off-grid position updates.

\subsection{Continuous Joint Design Formulation}
We now relax the grid restriction and allow the \(S\) active ports to
take continuous positions within the same aperture \(W_x\times W_y\).
The positions of the \(S\) active ports are collected in
\begin{equation}
\mathbf R=[\mathbf r_1,\ldots,\mathbf r_S]^\trans\in\R^{S\times2},
\label{eq:continuous_positions}
\end{equation}
where \(\mathbf r_s=[x_s,y_s]^\trans\) denotes the position of the
\(s\)-th active port. Each port coordinate is constrained to lie within
the physical aperture:
\begin{equation}
0\le x_s\le W_x,~0\le y_s\le W_y,~s=1,\ldots,S.
\label{eq:aperture_constraint}
\end{equation}
The active ports satisfy a prescribed minimum inter-port distance:
\begin{equation}
\|\mathbf r_s-\mathbf r_t\|_2\ge d_{\min},~
\forall s\ne t.
\label{eq:continuous_spacing}
\end{equation}

For a given placement \(\mathbf R\), the position-dependent steering
matrix \(\mathbf D(\mathbf R)\in\C^{Z\times S}\) (the symbol
\(\mathbf D\) is reused here with dimensions adapted to the active-port
set) has entries
\begin{equation}
[\mathbf D(\mathbf R)]_{z,s}
=\exp[-j2\pi(x_su_z+y_sv_z)].
\label{eq:continuous_dictionary}
\end{equation}
Let \(\mathbf w=[w_1,\ldots,w_S]^\trans\in\C^{S\times1}\) collect the
complex beamforming weights of the active ports. For
\(w_s=|w_s|e^{j\psi_s}\), the magnitude \(|w_s|\) and phase
\(\psi_s=\angle w_s\) specify the excitation amplitude and phase of port
\(s\), respectively. The corresponding beam is
\begin{equation}
\mathbf y
=\mathbf D(\mathbf R)\mathbf w.
\label{eq:continuous_beam}
\end{equation}
Continuous port-position optimization and beam synthesis are jointly
formulated as
\begin{equation}
\begin{gathered}
\min_{\mathbf R,\mathbf w}~
    \|\mathbf g-\mathbf D(\mathbf R)\mathbf w\|_2^2\\
\mathrm{s.t.}~
    0\le x_s\le W_x,~0\le y_s\le W_y,~
    s=1,\ldots,S,\\
\|\mathbf r_s-\mathbf r_t\|_2\ge d_{\min},~
s\ne t.
\end{gathered}
\label{eq:continuous_problem}
\end{equation}
Unlike the discrete formulation, this model allows each active port to take any continuous position satisfying the aperture and spacing constraints, thereby removing the candidate-grid restriction. Directly solving \eqref{eq:continuous_problem} is challenging because the steering matrix depends nonlinearly on the port coordinates and the spacing constraints define a nonconvex feasible set. We therefore retain these constraints but replace the squared-error objective with the performance-oriented loss in Section~\ref{sec:continuous-adam}. The OMP-selected port coordinates and corresponding least-squares weights initialize the projected Adam refinement.

\section{Proposed OMP--Adam Beamforming Framework}\label{sec:problem}

As summarized in Fig.~\ref{fig:framework-overview}, the proposed method
first converts the prescribed amplitude mask into a complex target
through finite-support phase retrieval and then uses
OMP to select \(S\) candidate ports and compute their least-squares
weights. In the second stage, this discrete solution initializes the
joint Adam refinement of the continuous port coordinates and complex
weights. The following subsections establish the Fourier relation used
by phase retrieval and detail the two optimization stages.

\subsection{Array--Beam Fourier Relation}
Under the adopted far-field model, the array factor of a planar aperture
is related to the spatial distribution of its complex beamforming weights through a
two-dimensional spatial Fourier transform. For the discrete candidate
grid in \eqref{eq:discrete_grid}, let
\(\mathbf W\in\C^{M_c\times N_c}\) denote the grid weight matrix,
where \([\mathbf W]_{i,j}\) is the complex beamforming weight assigned to the
candidate position \((x_i,y_j)\). For a realizable sparse grid
solution, entries corresponding to inactive ports are zero.
Here, \(i\) and \(j\) index the aperture samples along the \(x\)- and
\(y\)-axes, respectively. The indices \(p\) and \(q\) identify samples on the two-dimensional
beam-domain grid. The index \(z\) identifies the vectorized samples retained
inside the visible disk.
The column-major vectorization of \(\mathbf W\) gives the weight vector
\(\mathbf w\) introduced in the discrete formulation. The array factor
evaluated at \((u,v)\) can then be written as
\begin{equation}
    Y(u,v)=\sum_{i=1}^{M_c}\sum_{j=1}^{N_c}
    [\mathbf W]_{i,j}e^{-j2\pi(x_i u+y_j v)}.
    \label{eq:angle-domain-beam}
\end{equation}
Here, \((u,v)\) and \((\theta,\varphi)\) are related by
\eqref{eq:uv_mapping}. Let
\(\mathbf Y\in\C^{P\times Q}\) collect the sampled array-factor values,
with \([\mathbf Y]_{p,q}=Y(u_p,v_q)\). After restriction to the visible
disk and vectorization, these samples yield the beam vector \(\mathbf y\)
in \eqref{eq:discrete_beam}. When \((u,v)\) is sampled on the reciprocal grid
associated with the uniformly spaced aperture samples, the summation in
\eqref{eq:angle-domain-beam} reduces, up to indexing, centering, and
normalization conventions, to a two-dimensional discrete Fourier
transform (DFT). Conversely, an inverse DFT, implemented efficiently by
the inverse FFT (IFFT), maps a complex beam-domain
matrix \(\mathbf G\in\C^{P\times Q}\) to a sampled aperture distribution
of the form
\begin{equation}
    [\mathbf W]_{i,j}\propto\sum_{p=1}^{P}\sum_{q=1}^{Q}
    [\mathbf G]_{p,q}
    e^{j2\pi d_{\mathrm{cand}}[(i-1)u_p+(j-1)v_q]}.
    \label{eq:inverse-sum}
\end{equation}
The proportionality absorbs the inverse-transform normalization. This
Fourier relation connects uniformly sampled aperture weights with their
reciprocal beam-domain samples. For phase retrieval, we use an auxiliary
\(L\times L\) aperture support with spacing \(d_{\mathrm{PR}}\).
Its reciprocal grid defines the \(P\times Q\) FFT workspace. The candidate
port dictionary uses spacing \(d_{\mathrm{cand}}\) and is evaluated directly
on the retained visible-domain samples. OMP and least squares determine
the active-port support and its complex weights on this candidate grid.

\subsection{Discrete Stage: Phase Retrieval and Port Selection}
\subsubsection{OMP Port Selection}
Recall the sparse approximation problem in \eqref{eq:discrete_problem},
where \(\mathbf D\) is the antenna steering dictionary,
\(\mathbf g\) is the target beam, and \(\mathbf w\) contains
the complex beamforming weights. Our objective is to select \(S\)
suitable positions from \(N_p\) candidates to satisfy the
\(\ell_0\) constraint. This can be viewed as a sparse signal recovery problem \cite{sp1,sp2,sp3,sp4}, where \(\mathbf g\) is the observed signal, \(\mathbf D\)
is the dictionary, \(\mathbf w\) is the sparse
coefficient vector, and \(S\) is the sparsity. To this end, this paper
uses standard OMP for antenna
port selection. We first present the OMP update for a given complex
target \(\mathbf g\). The subsequent subsection explains how this target
is constructed from an amplitude-only specification before OMP is run.
Let \(\mathbf d_l=[\mathbf D]_{:,l}\) denote the \(l\)-th dictionary
column. The dictionary columns are normalized as
\begin{equation}
    \widetilde{\mathbf d}_l
    =\frac{\mathbf d_l}{\|\mathbf d_l\|_2}.
    \label{eq:normalized-dict}
\end{equation}
The normalized dictionary is
\(\widetilde{\mathbf D}=[\widetilde{\mathbf d}_1,\ldots,
\widetilde{\mathbf d}_{N_p}]\).
Set \(\mathbf g\) as the initial residual \(\mathbf e_0\) and let
\(\mathcal A_0=\emptyset\) be the support set. At iteration \(t\), the
atom with the largest absolute residual correlation is selected:
\begin{equation}
    l_t=\arg\max_{l\notin\mathcal A_{t-1}}
    \bigl|\widetilde{\mathbf d}_l^{\herm}\mathbf e_{t-1}\bigr|.
    \label{eq:omp-selection}
\end{equation}
The selected index is added to the support,
\(\mathcal A_t\leftarrow\mathcal A_{t-1}\cup\{l_t\}\).

The residual is updated via the standard OMP least-squares projection,
\begin{equation}
    \widehat{\mathbf{x}}_{\mathcal A_t}
    =[\widetilde{\mathbf D}]_{:,\mathcal A_t}^{\dagger}\mathbf g,
    \label{eq:omp-ls}
\end{equation}
\begin{equation}
    \mathbf e_t=\mathbf g
    -[\widetilde{\mathbf D}]_{:,\mathcal A_t}
    \widehat{\mathbf{x}}_{\mathcal A_t}.
    \label{eq:omp-residual}
\end{equation}
The iteration is repeated until \(S\) atoms have been selected.
After the final support \(\mathcal A_S\) is obtained, the actual
beamforming weights are recomputed with the unnormalized selected
dictionary,
\begin{equation}
    \widehat{\mathbf w}_{\mathcal A_S}
    =[\mathbf D]_{:,\mathcal A_S}^{\dagger}\mathbf g,
    \label{eq:final-weight}
\end{equation}
and the corresponding synthesized beam is
\begin{equation}
    \widehat{\mathbf y}
    =[\mathbf D]_{:,\mathcal A_S}
    \widehat{\mathbf w}_{\mathcal A_S}.
    \label{eq:final-beam}
\end{equation}
The final weights are intentionally recomputed with the unnormalized
selected dictionary in \eqref{eq:final-weight}, which restores the
physical scaling that is removed only for fair atom selection during
OMP. The selected positions and the corresponding least-squares weights
also provide the initial point for the subsequent continuous refinement
stage.
The complete OMP-based port-selection procedure is summarized in
Algorithm~\ref{alg:omp}.

\begin{algorithm}[!t]
\caption{OMP-Based Port Selection for the FAA}
\label{alg:omp}
\begin{algorithmic}[1]
\STATE \textbf{Input:} Dictionary \(\mathbf D\), complex
target beam \(\mathbf g\), and number of active ports \(S\).
\STATE \textbf{Output:} Active support set \(\mathcal A_S\).
\STATE Normalize columns of \(\mathbf D\) to obtain
\(\widetilde{\mathbf D}\).
\STATE Initialize \(\mathbf e_0\leftarrow\mathbf g\),
\(\mathcal A_0\leftarrow\emptyset\).
\FOR{\(t=1,2,\ldots,S\)}
\STATE Select \(l_t\) as the atom not yet selected with the maximum
absolute residual correlation according to \eqref{eq:omp-selection}.
\STATE Update \(\mathcal A_t\leftarrow\mathcal A_{t-1}\cup\{l_t\}\).
\STATE Solve the least-squares problem in \eqref{eq:omp-ls}.
\STATE Update the residual according to \eqref{eq:omp-residual}.
\ENDFOR
\STATE \textbf{Return} \(\mathcal A_S\).
\end{algorithmic}
\end{algorithm}

With a direct least-squares solve at every iteration,
Algorithm~\ref{alg:omp} has complexity
\(\mathcal O(SZN_p+ZS^3)\). The two terms arise from residual
correlation over the candidate dictionary and repeated least-squares
fitting, respectively.

\subsubsection{Phase Retrieval}
Practical beam specifications commonly prescribe an amplitude mask but
leave the phase unconstrained. Directly assigning an arbitrary phase can
produce a complex target that is difficult to approximate with a finite
aperture. We therefore use the Gerchberg--Saxton alternating-projection
method~\cite{gs1972} to seek a phase that is more
compatible with the aperture support. Let
\(\mathbf G_{\mathrm{am}}\in\R_+^{P\times Q}\) be the
discretized target amplitude, i.e.,
\([\mathbf G_{\mathrm{am}}]_{p,q}=G_{\mathrm{am}}(u_p,v_q)\). The
complex target is parameterized entrywise by an unknown phase matrix
\(\boldsymbol\Phi\in\R^{P\times Q}\):
\begin{equation}
    [\mathbf G]_{p,q}
    =[\mathbf G_{\mathrm{am}}]_{p,q}
    e^{j[\boldsymbol\Phi]_{p,q}},
    \label{eq:phase-target}
\end{equation}
for \((p,q)\in\mathcal I_{\mathcal V}\). The entries outside
\(\mathcal I_{\mathcal V}\) serve as auxiliary FFT variables.
The beam-domain matrix \(\mathbf G\) and the auxiliary aperture
distribution form the two domains of the alternating projections.
The FFT and IFFT connect these domains to estimate a target phase
compatible with the finite-aperture support.

We initialize the iteration with a broadside-referenced linear phase.
Let \((p_0,q_0)\) denote the grid indices corresponding to \(u=v=0\),
and set \([\boldsymbol\Phi^{(0)}]_{p,q}=-k[(p-p_0)+(q-q_0)]\),
where \(k\) is the phase slope. The initial matrix is
\begin{equation}
    [\mathbf G^{(0)}]_{p,q}=
    \begin{cases}
    [\mathbf G_{\mathrm{am}}]_{p,q}e^{j[\boldsymbol\Phi^{(0)}]_{p,q}},
        &(p,q)\in\mathcal I_{\mathcal V},\\
    0, &(p,q)\notin\mathcal I_{\mathcal V}.
    \end{cases}
    \label{eq:initial-complex-target}
\end{equation}
At iteration \(t\), the inverse Fourier transform maps the current beam
to the array domain:
\begin{equation}
    \mathbf A^{(t)}=\mathcal F^{-1}\{\mathbf G^{(t)}\}.
    \label{eq:ifft-step}
\end{equation}
Under the FFT indexing convention used in the implementation, we retain
the upper-left \(L\times L\) block of \(\mathbf A^{(t)}\) and set all
other samples to zero, yielding \(\widetilde{\mathbf A}^{(t)}\) and
imposing the finite-aperture support. The truncated distribution is then
mapped back to the beam domain:
\begin{equation}
    \widetilde{\mathbf G}^{(t)}
    =\mathcal F\{\widetilde{\mathbf A}^{(t)}\}.
    \label{eq:fft-step}
\end{equation}
The intermediate field \(\widetilde{\mathbf G}^{(t)}\) is generated by
the finite-support aperture distribution. The next beam-domain projection
restores the prescribed amplitude inside the visible disk and retains
the intermediate field outside it:
\begin{equation}
    [\mathbf G^{(t+1)}]_{p,q}=
    \begin{cases}
    [\mathbf G_{\mathrm{am}}]_{p,q}e^{j\angle[\widetilde{\mathbf G}^{(t)}]_{p,q}},
        &(p,q)\in\mathcal I_{\mathcal V},\\
    [\widetilde{\mathbf G}^{(t)}]_{p,q},
        &(p,q)\notin\mathcal I_{\mathcal V}.
    \end{cases}
    \label{eq:phase-update}
\end{equation}
After \(T_{\mathrm{PR}}\) alternating projections, the visible-domain
entries preserve the specified amplitude and carry the retrieved phase.
These entries are vectorized to form the OMP target \(\mathbf g\).
The FFT-based alternating-projection procedure is summarized in
Algorithm~\ref{alg:phase}.

\begin{algorithm}[!t]
\caption{FFT-Based Phase Retrieval}
\label{alg:phase}
\begin{algorithmic}[1]
\STATE \textbf{Input:} Target amplitude \(\mathbf G_{\mathrm{am}}\), visible-domain
indices \(\mathcal I_{\mathcal V}\), \(L\times L\) aperture support,
number of iterations \(T_{\mathrm{PR}}\), and phase slope \(k\).
\STATE \textbf{Output:} Complex target beam
\(\mathbf G^{(T_{\mathrm{PR}})}\).
\STATE Initialize the broadside-referenced linear phase and construct \(\mathbf G^{(0)}\)
according to \eqref{eq:initial-complex-target}.
\FOR{\(t=0,1,\ldots,T_{\mathrm{PR}}-1\)}
\STATE \(\mathbf A^{(t)}\leftarrow\mathcal F^{-1}\{\mathbf G^{(t)}\}\).
\STATE Retain the upper-left \(L\times L\) block of \(\mathbf A^{(t)}\)
and set all other entries to zero.
\STATE \(\widetilde{\mathbf G}^{(t)}
\leftarrow\mathcal F\{\widetilde{\mathbf A}^{(t)}\}\).
\STATE Update \(\mathbf G^{(t+1)}\) according to
\eqref{eq:phase-update}.
\ENDFOR
\end{algorithmic}
\end{algorithm}

Each iteration of Algorithm~\ref{alg:phase} performs one
two-dimensional IFFT and one two-dimensional FFT over the
\(P\times Q\) grid. Its complexity is therefore
\(\mathcal O(T_{\mathrm{PR}}PQ\log(PQ))\).

Within the visible disk, phase retrieval preserves the prescribed
amplitude and updates the target phase. All configurations therefore
share the same target amplitude and mainlobe region. The GS-assisted
configurations use the retrieved phase for the complex OMP target.
The target amplitude used in the simulations is detailed in
Section~\ref{sec:simu-setup}.

\subsection{Continuous Stage: Projected Adam Joint Refinement}
\label{sec:continuous-adam}

The OMP stage provides a physically meaningful but grid-constrained
initial solution. To exploit the continuous aperture, projected Adam
jointly refines the selected port coordinates and their complex
beamforming weights.

\subsubsection{Optimization Variables and Initialization}
Using the position matrix \(\mathbf R\) and steering matrix
\(\mathbf D(\mathbf R)\) defined in
\eqref{eq:continuous_positions}--\eqref{eq:continuous_dictionary}, we
write the vector of complex beamforming weights as
\(\mathbf w=\mathbf w_r+j\mathbf w_i\), where
\(\mathbf w_r,\mathbf w_i\in\R^{S\times1}\). The continuous Adam stage
therefore jointly optimizes the three real-valued variable blocks
\(\mathbf R\), \(\mathbf w_r\), and \(\mathbf w_i\).

The OMP solution initializes the continuous positions as
\(\mathbf R^{(0)}=\mathbf R_{\mathrm{OMP}}\).
With \(\mathbf D(\mathbf R_{\mathrm{OMP}})\) denoting the corresponding
unnormalized steering matrix, least squares gives the initial complex
beamforming weights:
\begin{equation}
    \mathbf w^{(0)}
    =\mathbf D(\mathbf R_{\mathrm{OMP}})^\dagger\mathbf g.
    \label{eq:adam-weight-init}
\end{equation}
Adam then jointly updates the real and imaginary parts of the weights
and the port coordinates.

\subsubsection{Loss Function}
The synthesized array response is
\(\mathbf y=\mathbf D(\mathbf R)(\mathbf w_r+j\mathbf w_i)\),
with squared magnitude \(q_z=|y_z|^2\) at sample \(z\).
For \(N_{\mathrm{beam}}\) desired beams, let
\(\mathcal B_b\subseteq\{1,\ldots,Z\}\) denote the \(b\)-th mainlobe
sample set and \(\mathcal B=\bigcup_b\mathcal B_b\) their union.
The mainlobe energy fraction is evaluated on the uniform \((u,v)\)
grid within the visible disk:
\begin{equation}
    \eta_{\mathcal B}
    =\frac{\sum_{z\in\mathcal B}q_z}{\sum_{z=1}^{Z}q_z}.
    \label{eq:mainlobe-energy-fraction}
\end{equation}
It measures the fraction of the sampled response energy contained in
the prescribed mainlobe region. We express the ratio between the average
mainlobe response energy and the average visible-domain response energy
as the mainlobe concentration,
\begin{equation}
    C_{\mathcal B}
    =\frac{|\mathcal B|^{-1}\sum_{z\in\mathcal B}q_z}
    {Z^{-1}\sum_{z=1}^{Z}q_z}
    =\frac{Z}{|\mathcal B|}\eta_{\mathcal B},
    \label{eq:mainlobe-concentration}
\end{equation}
and use its logarithmic form in the optimization:
\begin{equation}
    c_{\mathcal B}=10\log_{10}C_{\mathcal B}.
    \label{eq:concentration-db}
\end{equation}
For a prescribed mainlobe region, maximizing \(c_{\mathcal B}\)
also maximizes \(\eta_{\mathcal B}\). Both quantities are invariant
to a common amplitude scaling of \(\mathbf y\).

The concentration of each beam is
\begin{equation}
    c_b=10\log_{10}\!\left[
    \frac{|\mathcal B_b|^{-1}\sum_{z\in\mathcal B_b}q_z}
    {Z^{-1}\sum_{z=1}^{Z}q_z}\right].
    \label{eq:beam-concentration-db}
\end{equation}
We balance the beams through the variance of these values:
\begin{equation}
    \mathcal L_{\mathrm{bal}}
    =\frac{1}{N_{\mathrm{beam}}}\sum_{b=1}^{N_{\mathrm{beam}}}
    (c_b-\bar c)^2,~
    \bar c=\frac{1}{N_{\mathrm{beam}}}\sum_{b=1}^{N_{\mathrm{beam}}}c_b.
    \label{eq:balance-loss}
\end{equation}
The continuous refinement uses the loss
\begin{equation}
    \mathcal L(\mathbf R,\mathbf w_r,\mathbf w_i)
    =-c_{\mathcal B}+\beta\mathcal L_{\mathrm{bal}},
    \label{eq:total-loss}
\end{equation}
Here \(\beta\) controls the beam-balance weight. Spacing repair
enforces the minimum interelement distance after every position update.

Fitting \(\mathbf g\) supplies the initial port layout and weights,
while continuous refinement promotes energy concentration within
the prescribed mainlobe regions and balance among beams.
PSLL provides a complementary evaluation of the strongest sidelobe.

\subsubsection{Adam Optimization}
At iteration \(t\), the loss in \eqref{eq:total-loss} is evaluated at the
current state \((\mathbf R_t,\mathbf w_{r,t},\mathbf w_{i,t})\).
Automatic differentiation then returns all three gradients in one pass:
\begin{equation}
\begin{aligned}
    \mathbf H_{R,t}
    &=
    \nabla_{\mathbf R}
    \mathcal L(\mathbf R_t,\mathbf w_{r,t},\mathbf w_{i,t}),\\
    \mathbf h_{r,t}
    &=
    \nabla_{\mathbf w_r}
    \mathcal L(\mathbf R_t,\mathbf w_{r,t},\mathbf w_{i,t}),\\
    \mathbf h_{i,t}
    &=
    \nabla_{\mathbf w_i}
    \mathcal L(\mathbf R_t,\mathbf w_{r,t},\mathbf w_{i,t}).
\end{aligned}
    \label{eq:joint-gradients}
\end{equation}
The three variable blocks share the same loss but maintain independent
Adam first- and second-moment states. Thus,
\begin{equation}
\begin{aligned}
    \mathbf R_{t+1/2}
    &=
    \operatorname{Adam}_{R}
    (\mathbf R_t,\mathbf H_{R,t};\eta_{R,t}),\\
    \mathbf w_{r,t+1}
    &=
    \operatorname{Adam}_{r}
    (\mathbf w_{r,t},\mathbf h_{r,t};\eta_{w,t}),\\
    \mathbf w_{i,t+1}
    &=
    \operatorname{Adam}_{i}
    (\mathbf w_{i,t},\mathbf h_{i,t};\eta_{w,t}).
\end{aligned}
    \label{eq:adam-block-update}
\end{equation}
The gradients in \eqref{eq:joint-gradients} are all evaluated at the
same state, and Adam jointly updates all variables.
\begin{algorithm}[t!]
	\caption{Projected Adam Refinement}
	\label{alg:adam}
	\begin{algorithmic}[1]
		\STATE \textbf{Input:} \(\mathbf g\), \(\mathbf R_{\mathrm{OMP}}\),
		\(\{(u_z,v_z)\}_{z=1}^{Z}\), \(\{\mathcal B_b\}\),
		\(T_{\mathrm{Adam}}\), \(d_{\min}\), \(\beta\), \(\Delta_{\max}\).
		\STATE \textbf{Output:} \(\mathbf R^\star\), \(\mathbf w^\star\).
		\STATE \(\mathbf R\leftarrow\mathbf R_{\mathrm{OMP}}\),
		\(\mathbf w^{(0)}\leftarrow\mathbf D(\mathbf R)^\dagger\mathbf g\).
		\STATE \((\mathbf w_r,\mathbf w_i)\leftarrow
		(\Real\{\mathbf w^{(0)}\},\operatorname{Im}\{\mathbf w^{(0)}\})\).
		\STATE Initialize the Adam moments and the best feasible state.
		\FOR{\(t=1,2,\ldots,T_{\mathrm{Adam}}\)}
		\STATE Evaluate \(\mathcal L\) and \((\mathbf H_R,\mathbf h_r,\mathbf h_i)\)
		via \eqref{eq:total-loss} and \eqref{eq:joint-gradients}.
		\STATE \(\widehat{\mathbf R}\leftarrow
		\operatorname{Adam}_{R}(\mathbf R,\mathbf H_R,\eta_{R,t})\).
		\STATE \((\mathbf R,\zeta_t)\leftarrow
		\mathcal R_{d_{\min}}(\Pi_{\mathcal A}(\widehat{\mathbf R}))\).
		\STATE \(\mathbf w_\chi\leftarrow
		\operatorname{Adam}_\chi(\mathbf w_\chi,\mathbf h_\chi,\eta_{w,t})\),
		\(\chi\in\{r,i\}\).
		\STATE Rescale \(\mathbf w_r+j\mathbf w_i\) to
		\(\|\mathbf w^{(0)}\|_2\), and evaluate \(c_{\mathcal B}\) and
		\(\Delta_c=\max_bc_b-\min_bc_b\).
		\IF{finite, \(\zeta_t=1\),
			\(\Delta_c\le\Delta_{\max}\), and
			\(c_{\mathcal B}>c_{\mathrm{best}}\)}
		\STATE \((\mathbf R_{\mathrm{best}},\mathbf w_{\mathrm{best}},
		c_{\mathrm{best}})\leftarrow
		(\mathbf R,\mathbf w_r+j\mathbf w_i,c_{\mathcal B})\).
		\ENDIF
		\ENDFOR
		\STATE \((\mathbf R^\star,\zeta^\star)\leftarrow
		\mathcal R_{d_{\min}}(\mathbf R_{\mathrm{best}})\),
		\(\mathbf w^\star\leftarrow\mathbf w_{\mathrm{best}}\).
		\IF{\(\zeta^\star=0\)}
		\STATE Report failure of the final minimum-spacing check.
		\ENDIF
		\STATE \textbf{Return} \(\mathbf R^\star\), \(\mathbf w^\star\).
	\end{algorithmic}
\end{algorithm}

After each position update, the coordinates are first projected
elementwise onto the aperture \(\mathcal A=[0,W_x]\times[0,W_y]\) and
then passed to a minimum-spacing repair operator:
\begin{equation}
\begin{aligned}
    \widetilde{\mathbf R}_{t+1}
    &=\Pi_{\mathcal A}(\mathbf R_{t+1/2}),\\
    (\mathbf R_{t+1},\zeta_t)
    &=\mathcal R_{d_{\min}}
    (\widetilde{\mathbf R}_{t+1}),
\end{aligned}
    \label{eq:clamp}
\end{equation}
where \(\mathcal R_{d_{\min}}(\cdot)\) denotes the iterative
minimum-spacing repair specified in Algorithm~\ref{alg:project}, and
\(\zeta_t\) indicates whether the repaired placement satisfies the
spacing constraint. This repair resolves pairwise spacing violations
and preserves the Adam moment states. After each weight update, the
vector of complex beamforming weights is rescaled to preserve its initial norm
\(\|\mathbf w^{(0)}\|_2\), which removes an uninformative scale degree of
freedom because the loss in \eqref{eq:total-loss} depends only on power
ratios.

The position learning rate follows a cosine schedule, while the weight
learning rate uses a short linear warm-up followed by cosine decay. The
cosine-decay portion is
\begin{equation}
    \eta_{\chi,t}
    =\eta_{\chi,\min}
    +\frac{1}{2}(\eta_{\chi,\max}-\eta_{\chi,\min})
    \bigl(1+\cos(\pi\tau_{\chi,t})\bigr),
    \label{eq:cosine-lr}
\end{equation}
where \(\chi\in\{R,w\}\) and \(\tau_{\chi,t}\) increases from zero to
one over the corresponding decay interval. For the weight block, this
interval begins after the warm-up period. The two blocks use independently
specified maximum and minimum learning rates. During optimization, a checkpoint is
eligible if all quantities are finite, its minimum port spacing is at
least \(d_{\min}\), and its maximum inter-beam concentration gap does not exceed a
prescribed threshold. Among all eligible checkpoints, the Adam
refinement returns the one with the largest \(c_{\mathcal B}\). Each
target beam configuration is optimized independently. If no Adam iterate
is eligible, the feasible OMP initialization is retained.
The complete projected Adam refinement is summarized in
Algorithm~\ref{alg:adam}.

The initial least-squares fit in Algorithm~\ref{alg:adam} costs
\(\mathcal O(ZS^2)\). Each Adam iteration requires
\(\mathcal O(ZS)\) operations for field and gradient evaluation and
\(\mathcal O(T_{\mathrm{proj}}S^2)\) operations for spacing repair,
giving a total complexity of
\(\mathcal O(ZS^2+T_{\mathrm{Adam}}[ZS+T_{\mathrm{proj}}S^2])\).

The minimum-spacing repair is invoked after every position update, with a
final invocation verifying the spacing feasibility. It applies iterative
pairwise repulsion: for every violating pair \((i,j)\), both ports are
pushed apart along their connecting line by half the violation distance,
and the result is clipped back to the physical aperture. The procedure
terminates when a complete sweep finds no violation or when the maximum
number of repair sweeps is reached, after which feasibility is reported.
The detailed procedure is given in Algorithm~\ref{alg:project}.

\begin{algorithm}[t!]
\caption{Iterative Minimum-Spacing Repair}
\label{alg:project}
\begin{algorithmic}[1]
\STATE \textbf{Input:} Port positions \(\mathbf R\in\R^{S\times2}\),
minimum spacing \(d_{\min}\), aperture dimensions \(W_x,W_y\), and maximum
number of repair sweeps \(T_{\mathrm{proj}}\).
\STATE \textbf{Output:} Repaired positions \(\mathbf R^\star\) and
feasibility indicator \(\zeta\).
\STATE \(\mathbf P\leftarrow\Pi_{\mathcal A}(\mathbf R)\), where
\(\mathcal A=[0,W_x]\times[0,W_y]\).
\STATE Let \(\mathbf P=[\mathbf p_1,\ldots,\mathbf p_S]^\trans\).
\FOR{\(m=1,2,\ldots,T_{\mathrm{proj}}\)}
\STATE \(\mathrm{moved}\leftarrow\mathrm{false}\).
\FOR{each pair \(i<j\)}
\STATE \(\mathbf d_{ij}\leftarrow\mathbf p_j-\mathbf p_i\),
\(d\leftarrow\|\mathbf d_{ij}\|_2\).
\IF{\(d<d_{\min}\)}
\STATE Set \(\mathbf q\leftarrow\mathbf d_{ij}/d\) if \(d>0\).
\STATE Otherwise, choose any unit vector \(\mathbf q\).
\STATE \(\delta\leftarrow(d_{\min}-d)/2\).
\STATE \(\mathbf p_i\leftarrow\mathbf p_i-\delta\mathbf q\),
\(\mathbf p_j\leftarrow\mathbf p_j+\delta\mathbf q\).
\STATE \(\mathrm{moved}\leftarrow\mathrm{true}\).
\ENDIF
\ENDFOR
\STATE \(\mathbf P\leftarrow\Pi_{\mathcal A}(\mathbf P)\).
\IF{\(\mathrm{moved}=\mathrm{false}\)}
\STATE \textbf{break}.
\ENDIF
\ENDFOR
\STATE \(\mathbf R^\star\leftarrow\mathbf P\).
\STATE \(\zeta\leftarrow
\bigl[\min_{1\le i<j\le S}\|\mathbf p_i-\mathbf p_j\|_2
\ge d_{\min}\bigr]\).
\STATE \textbf{Return} \(\mathbf R^\star\), \(\zeta\).
\end{algorithmic}
\end{algorithm}
\begin{table}[!t]
	\centering
	\caption{Simulation Parameters}
	\label{tab:simulation-parameters}
	\footnotesize
	\renewcommand{\arraystretch}{1.05}
	\begin{tabular}{@{}
			>{\raggedright\arraybackslash}p{0.38\columnwidth}
			>{\centering\arraybackslash}p{0.23\columnwidth}
			>{\raggedright\arraybackslash}p{0.27\columnwidth}@{}}
		\toprule
		Parameter & Symbol & Value\\
		\midrule
		Wavelength & \(\lambda\) & \(1\)\\
		Aperture & \(W_x\times W_y\) &
		\(4.5\lambda\times4.5\lambda\)\\
		Fixed array and active ports & \(L\times L,S\) & \(10\times10,100\)\\
		Port spacings & \(d_{\mathrm{fixed}},d_{\mathrm{cand}}\) &
		\(0.5\lambda,0.25\lambda\)\\
		Candidate grid size & \(N_p\) & \(19\times19=361\)\\
		Spatial-frequency grid & \(P\times Q\) & \(181\times181\)\\
		Beam centers & \(\theta_0,\Phi_B\) &
		\(45^\circ,\{30^\circ,60^\circ\}\)\\
		Beam-shape parameters & \(\Delta_\theta,\Delta_\varphi,\rho_{\mathrm{flat}}\) &
		\(8^\circ,8^\circ,0.45\)\\
		Phase-retrieval configuration & \(T_{\mathrm{PR}},k\) & \(100,0.1\)\\
		Mainlobe/support thresholds & \(G_{\mathrm{am}}\) & \(0.5,0.05\)\\
		Adam iterations & \(T_{\mathrm{Adam}}\) & \(300\)\\
		Position learning rates & \(\eta_{R,\max},\eta_{R,\min}\) &
		\(1.5\times10^{-3},10^{-5}\)\\
		Weight warm-up & \(T_{\mathrm{warm}}\) & \(20\)\\
		Weight learning rates & \(\eta_{w,\max},\eta_{w,\min}\) &
		\(10^{-3},10^{-6}\)\\
		Adam moment parameters & \(\beta_1,\beta_2,\epsilon\) &
		\(0.9,0.999,10^{-8}\)\\
		Beam-balance weight & \(\beta\) & \(0.02\)\\
		Minimum port spacing & \(d_{\min}\) & \(0.2\lambda\)\\
		Maximum repair sweeps & \(T_{\mathrm{proj}}\) & \(300\)\\
Inter-beam concentration gap limit & \(\Delta_{\max}\) & \(0.1\)~dB\\
		\bottomrule
	\end{tabular}
\end{table}
\begin{table}[!t]
	\centering
	\caption{Asymptotic Computational Complexity}
	\label{tab:complexity}
	\scriptsize
	\setlength{\tabcolsep}{2pt}
	\renewcommand{\arraystretch}{1.12}
	\begin{tabular}{@{}
			>{\raggedright\arraybackslash}p{0.21\columnwidth}
			>{\raggedright\arraybackslash}p{0.27\columnwidth}
			>{\centering\arraybackslash}p{0.44\columnwidth}@{}}
		\toprule
		Algorithm & Dominant operation & Time complexity\\
		\midrule
		OMP port selection & Correlations and repeated least squares &
		\(\mathcal O(SZN_p+ZS^3)\)\\
		FFT phase retrieval & Two-dimensional FFT and IFFT &
		\(\mathcal O(T_{\mathrm{PR}}PQ\log(PQ))\)\\
		Projected Adam & Initialization, gradients, and repair &
		\(\mathcal O(ZS^2+T_{\mathrm{Adam}}[ZS+T_{\mathrm{proj}}S^2])\)\\
		Spacing repair & Pairwise spacing checks &
		\(\mathcal O(T_{\mathrm{proj}}S^2)\)\\
		\bottomrule
	\end{tabular}
\end{table}

Each repair sweep in Algorithm~\ref{alg:project} examines all
\(S(S-1)/2\) port pairs. Its complexity is therefore
\(\mathcal O(T_{\mathrm{proj}}S^2)\). The dominant operations of all four
algorithms are summarized in Table~\ref{tab:complexity}.

\section{Numerical Results}\label{sec:numerical}

\subsection{Simulation Setup}\label{sec:simu-setup}

The main simulation parameters are summarized in
Table~\ref{tab:simulation-parameters}. The fixed and fluid arrays use the
same physical aperture and active-port count. The fixed baseline places
its ports on a uniform grid, whereas the fluid array selects them from a
denser candidate grid. This setting isolates the benefit of port-selection
freedom under the same aperture and hardware budget.

For convenience, GS is used only in the simulation labels to denote
phase retrieval.

\begin{figure}[!t]
	\centering
	\includegraphics[width=\columnwidth]{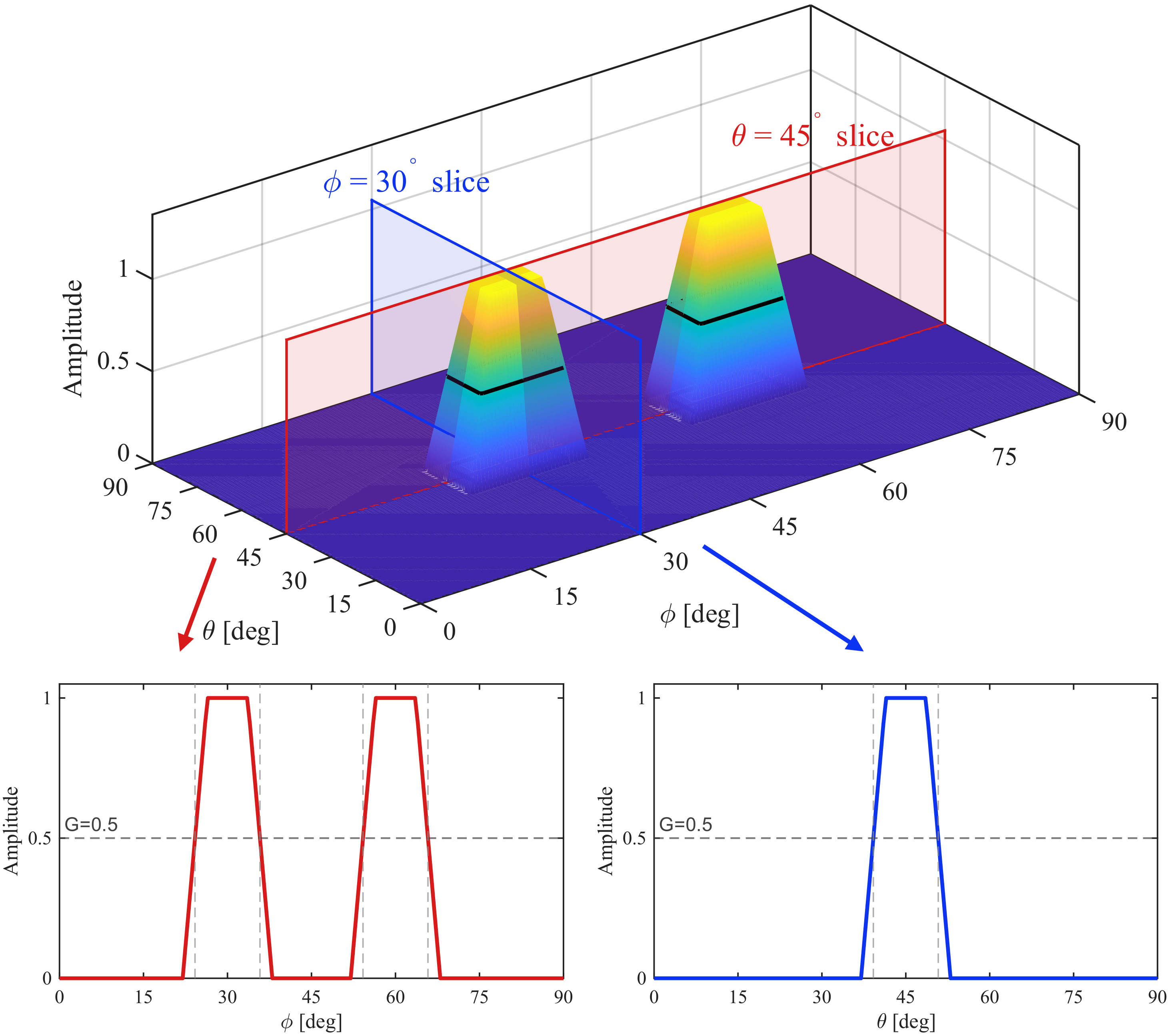}
	\caption{Desired amplitude pattern comprising two flat-top beams
		with linear roll-off.}
	\label{fig:desired-beam}
\end{figure}

The desired amplitude pattern comprises two flat-top beams with linear
roll-off. Both beams share the elevation center \(\theta_0\), while beam
\(b\) has azimuth center \(\varphi_b\), as specified in
Table~\ref{tab:simulation-parameters}. For beam \(b\), let
\(\rho_b(\theta,\varphi)\) denote the normalized maximum angular offset
of direction \((\theta,\varphi)\) from the beam center
\((\theta_0,\varphi_b)\), defined as
\begin{equation}
\rho_b(\theta,\varphi)
\triangleq
\max\left\{
\frac{|\theta-\theta_0|}{\Delta_\theta},
\frac{|\operatorname{wrap}(\varphi-\varphi_b)|}{\Delta_\varphi}
\right\},
\label{eq:normalized-angular-offset}
\end{equation}
where \(\operatorname{wrap}(\cdot)\) returns the principal azimuth
difference and thereby accounts for angular periodicity. The desired
amplitude is then
\begin{equation}
G_{\mathrm{am}}(\theta,\varphi)
=\max_{1\leq b\leq N_{\mathrm{beam}}}
\begin{cases}
1, & \rho_b\leq \rho_{\mathrm{flat}},\\
\dfrac{1-\rho_b}{1-\rho_{\mathrm{flat}}}, &
\rho_{\mathrm{flat}}<\rho_b\leq 1,\\
0, & \rho_b>1.
\end{cases}
\label{eq:target-amplitude}
\end{equation}
Here \(N_{\mathrm{beam}}=2\). The target peak amplitude is normalized to
one, and the pointwise maximum forms the union of the two beam templates.
This amplitude pattern defines the complex target for the discrete
least-squares fit and the mainlobe regions for continuous refinement.
Fig.~\ref{fig:desired-beam} illustrates the target pattern.
\begin{figure*}[t!]
	\centering
	\includegraphics[width=\textwidth]{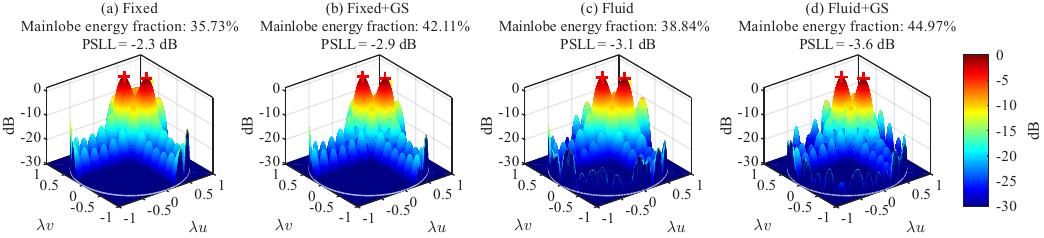}
	\caption{Discrete-stage comparison of three-dimensional beam patterns
		over the visible \(u,v\) disk under the same
		\(4.5\lambda\times4.5\lambda\) aperture and \(S=100\) active ports.
		From left to right, the panels show Fixed, Fixed+GS, Fluid, and Fluid+GS.
		Red crosses mark the mainlobe peaks. The respective mainlobe energy
		fractions are 35.73\%, 42.11\%,
		38.84\%, and 44.97\%, and the respective PSLL values are
		\(-2.30\), \(-2.89\), \(-3.14\), and \(-3.60\)~dB.}
	\label{fig:nr-discrete-4panel}
\end{figure*}
\begin{figure*}[!t]
	\centering
	\includegraphics[width=\textwidth]{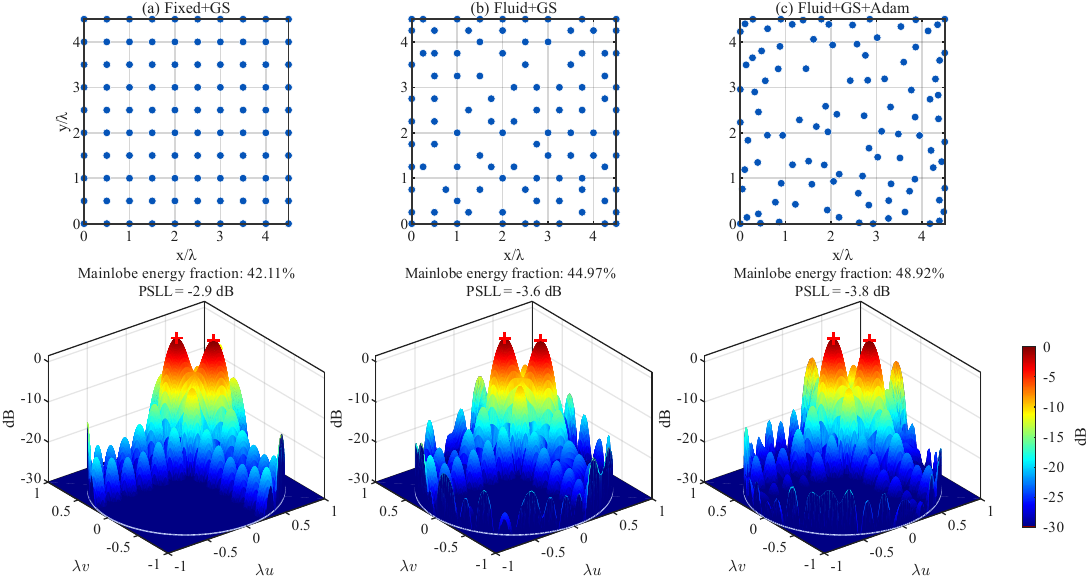}
	\caption{Continuous-stage comparison of active port placements and
		three-dimensional beam patterns. The columns show Fixed+GS, Fluid+GS, and
		Fluid+GS+Adam, respectively. The top row shows active port positions
		within the \(4.5\lambda\times4.5\lambda\) aperture, and the bottom row
		shows the corresponding beam patterns over the visible \(u,v\) disk.
		The respective
		mainlobe energy fractions are 42.11\%, 44.97\%, and 48.92\%, and the
		respective PSLL values are \(-2.89\), \(-3.60\), and \(-3.83\)~dB.}
	\label{fig:nr-continuous-3panel}
\end{figure*}
The selected shape parameters give each beam a flat-top half-width of
\(3.6^\circ\) and a linear transition half-width of \(4.4^\circ\).
All methods use the same spatial-frequency grid on the visible disk
\(\mathcal V=\{(u,v)\mid u^2+v^2\leq1\}\). The three-dimensional
patterns use normalized coordinates \((\lambda u,\lambda v)\), and the
angular cross sections span \([0,90^\circ]\).

\begin{figure}[!t]
	\centering
	\includegraphics[width=\columnwidth]{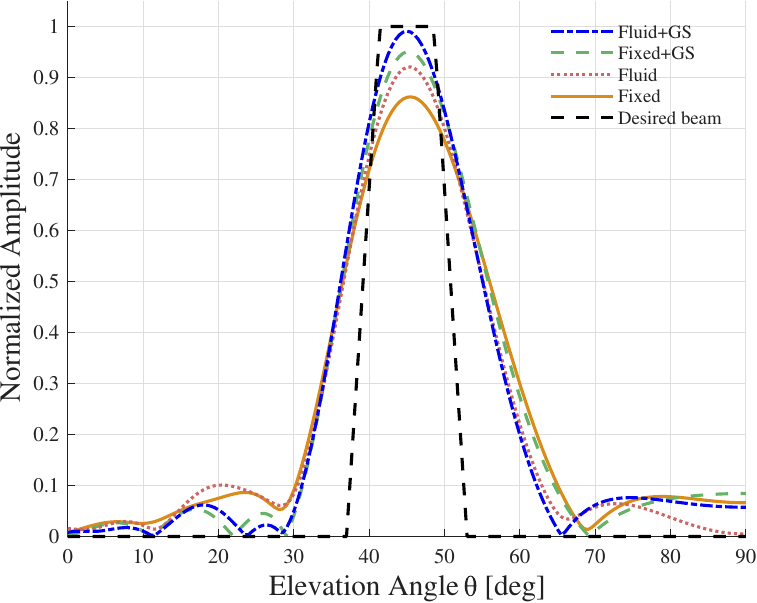}
	\caption{Discrete-stage normalized beam-amplitude cross section at
		\(\varphi=30^\circ\).}
	\label{fig:nr-discrete-cross-theta}
\end{figure}

The aperture, grid, and beam-shape parameters in
Table~\ref{tab:simulation-parameters} are shared by
Figs.~\ref{fig:nr-discrete-4panel}--\ref{fig:nr-cont-cross-theta}. The
phase-retrieval parameters apply to the GS-assisted configurations in
these figures, whereas the Adam and spacing-repair parameters apply only
to the Adam-refined results in
Figs.~\ref{fig:nr-continuous-3panel}--\ref{fig:nr-cont-cross-theta} and
Table~\ref{tab:init-comparison}.

The phase-retrieval iteration count and initial phase slope in
Table~\ref{tab:simulation-parameters} provide a common complex target for
all phase-retrieval-assisted methods. The auxiliary aperture support
uses \(d_{\mathrm{PR}}=0.5\lambda\).
Let \(\mathcal B=\{z\mid G_{\mathrm{am}}(u_z,v_z)\geq0.5\}\)
and \(\mathcal S=\{z\mid G_{\mathrm{am}}(u_z,v_z)\leq0.05\}\)
denote the mainlobe and sidelobe sample sets, respectively.
For \(\mathbf y=\mathbf D(\mathbf R)\mathbf w\), we report the
mainlobe energy fraction \(\eta_{\mathcal B}\) in
\eqref{eq:mainlobe-energy-fraction} as a percentage. The metrics use all
samples in the visible disk. The three-dimensional patterns use an
\(181\times181\) display grid, with each surface normalized by its displayed peak and
clipped at \(-30\)~dB. For the cross sections, each response
is first divided by \(\sqrt{\sum_{z=1}^{Z}|y_z|^2}\); all methods in a
comparison then share the maximum amplitude across both displayed cuts
as their display reference. The desired amplitude retains its unit peak.
The PSLL, normalized by the maximum visible-domain amplitude, is defined
as
\begin{equation}
\mathrm{PSLL}
=20\log_{10}\!\left(
\frac{\max_{z\in\mathcal S}|y_z|}
{\max_{1\leq k\leq Z}|y_k|}
\right).
\label{eq:setup-psll}
\end{equation}
The continuous stage is initialized by the Fluid+GS OMP solution. The position learning
rate follows cosine decay. The weight learning rate uses a short linear
warm-up followed by cosine decay. The weight norm is fixed to its initial
value because the loss depends only on energy ratios. The remaining Adam
and beam-balance parameters are listed in
Table~\ref{tab:simulation-parameters}.

Motivated by the minimum-spacing analyses for finite-aperture linear and
planar FAAs~\cite{faa_spacing_linear,faa_spacing_planar}, the iterative repair is
applied after every position update to enforce the minimum inter-port distance. A checkpoint is
eligible when all quantities are finite, the minimum spacing is feasible,
and the inter-beam concentration gap satisfies \(\Delta_c\le\Delta_{\max}\).
The eligible checkpoint with the highest \(c_{\mathcal B}\) is reported.

Using the problem dimensions and iteration counts in
Table~\ref{tab:simulation-parameters}, Table~\ref{tab:complexity}
summarizes the asymptotic cost of each proposed algorithm. The Projected
Adam entry includes the repair invoked after every position update,
whereas the spacing-repair entry gives the cost of one invocation.

\subsection{Discrete-Stage Comparison}
\label{sec:num-discrete}

We first compare four discrete-stage configurations before continuous
Adam refinement. The Fixed configuration uses the uniform half-wavelength
array and fits the prescribed flat-top amplitude target by plain least squares.
Fixed+GS uses the same fixed positions but fits the phase-retrieved complex
target. Fluid applies OMP to select \(S=100\) ports from the dense
candidate grid and fits the raw amplitude target. Fluid+GS combines OMP
selection with the phase-retrieved complex target. After the positions are
fixed, all four cases solve the beamforming weights by plain least squares.

For \(S=100\), the fixed baseline is a \(10\times10\) array densely packed
at a half-wavelength spacing. Its side length is therefore
\((\sqrt{S}-1)\lambda/2=4.5\lambda\). All four configurations use this
same \(4.5\lambda\times4.5\lambda\) physical aperture and the same number
of active ports. The Fluid configurations place the same \(S\)
ports by selecting their positions from the denser candidate grid.
\begin{figure}[!t]
	\centering
	\includegraphics[width=\columnwidth]{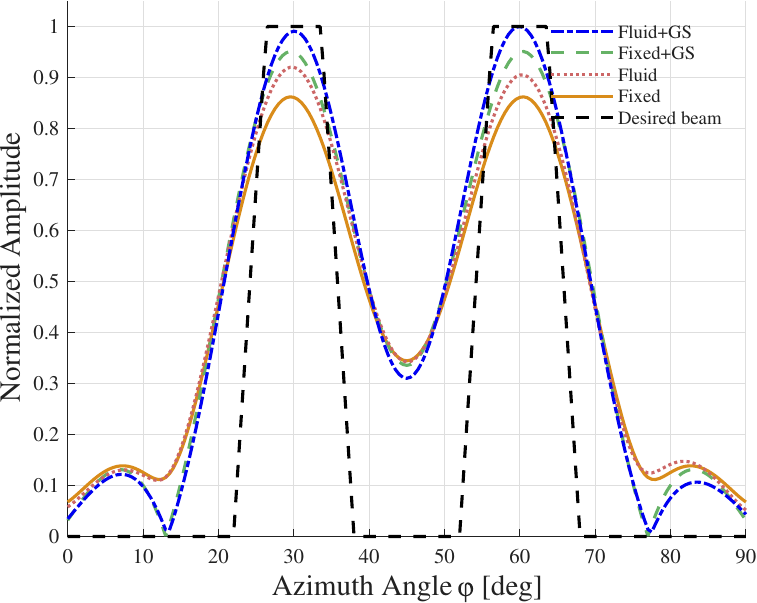}
	\caption{Discrete-stage normalized beam-amplitude cross section at
		\(\theta=45^\circ\).}
	\label{fig:nr-discrete-cross-phi}
\end{figure}

Fig.~\ref{fig:nr-discrete-4panel} compares the four three-dimensional beam
patterns over the visible disk. Phase retrieval increases the fixed-array mainlobe
energy fraction from 35.73\% to 42.11\%, an increase of 6.39 percentage
points. Fluid port selection raises this fraction from 35.73\% to
38.84\%, an increase of 3.11 percentage points. Combining both steps,
Fluid+GS reaches 44.97\%, exceeding Fixed+GS and Fixed by 2.86 and
9.24 percentage points, respectively. Its PSLL is \(-3.60\)~dB,
compared with \(-2.30\)~dB for Fixed.

Figs.~\ref{fig:nr-discrete-cross-phi} and~\ref{fig:nr-discrete-cross-theta}
show the corresponding normalized cross sections. At \(\theta=45^\circ\),
phase retrieval raises both main peaks, and Fluid+GS has the highest
peaks among the four discrete configurations. At \(\varphi=30^\circ\),
the combined design produces a more concentrated main beam and lower
nearby sidelobe shoulders.

\subsection{Continuous-Stage Comparison: Adam Refinement}
\label{sec:num-continuous}

We next activate the continuous refinement stage. The comparison focuses
on Fixed+GS, Fluid+GS, and Fluid+GS+Adam, as shown in
Fig.~\ref{fig:nr-continuous-3panel}. Fluid+GS+Adam starts from the OMP
positions and least-squares weights of Fluid+GS. Adam then jointly
refines the continuous port coordinates and the complex beamforming
weights using the concentration-and-balance loss in
\eqref{eq:total-loss}. Spacing repair follows every position update.

Adam refinement increases the mainlobe energy fraction from 44.97\%
to 48.92\%, adding 3.95 percentage points to the Fluid+GS result.
The complete pipeline exceeds Fixed+GS and Fixed by 6.80 and
13.19 percentage points, respectively. Adam also reduces the PSLL from
\(-3.60\)~dB to \(-3.83\)~dB. The final inter-beam concentration gap
is approximately \(0.08\)~dB.

To isolate the benefit of the OMP prior, we also run the same Adam
refinement from the uniform half-wavelength array, with its initial
weights fitted to the same phase-retrieved target by least squares. The
two runs use identical 300-iteration learning-rate schedules, refinement
parameters, and feasibility conditions.
Table~\ref{tab:init-comparison} reports the mainlobe energy fraction
of the best feasible checkpoint observed up to each sampled iteration.
Fluid+GS initialization provides a higher fraction at every listed
checkpoint. Its repeated values at iterations 20 and 50 correspond to
the retained initialization, since subsequent improvements at those
checkpoints have yet to satisfy the beam-balance condition. The advantage
is 2.22 percentage points at iteration 100 and 1.49 percentage points
at iteration 300. These results quantify the finite-budget benefit of
the OMP initialization under a common learning-rate schedule.

\begin{table}[!t]
\centering
\caption{Mainlobe Energy Fraction for Two Projected Adam Initializations}
\label{tab:init-comparison}
\begin{tabular}{ccc}
\toprule
Iteration & Fixed+GS start & Fluid+GS start \\
 & (\%) & (\%) \\
\midrule
0   & 42.11 & 44.97 \\
20  & 43.41 & 44.97 \\
50  & 44.51 & 44.97 \\
100 & 45.77 & 47.98 \\
200 & 47.18 & 48.80 \\
300 & 47.43 & 48.92 \\
\bottomrule
\end{tabular}
\end{table}
\begin{figure}[!t]
	\centering
	\includegraphics[width=\columnwidth]{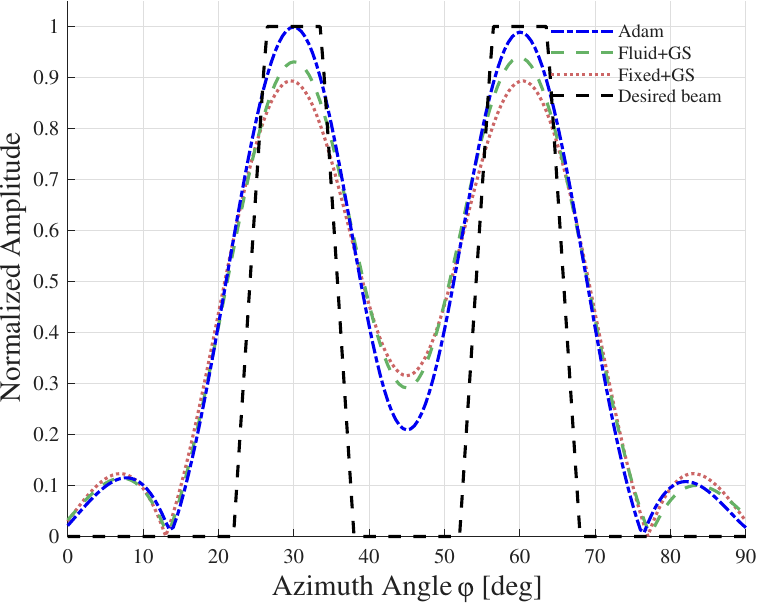}
	\caption{Continuous-stage normalized beam-amplitude cross section at
		\(\theta=45^\circ\).}
	\label{fig:nr-cont-cross-phi}
\end{figure}

\begin{figure}[!t]
	\centering
	\includegraphics[width=\columnwidth]{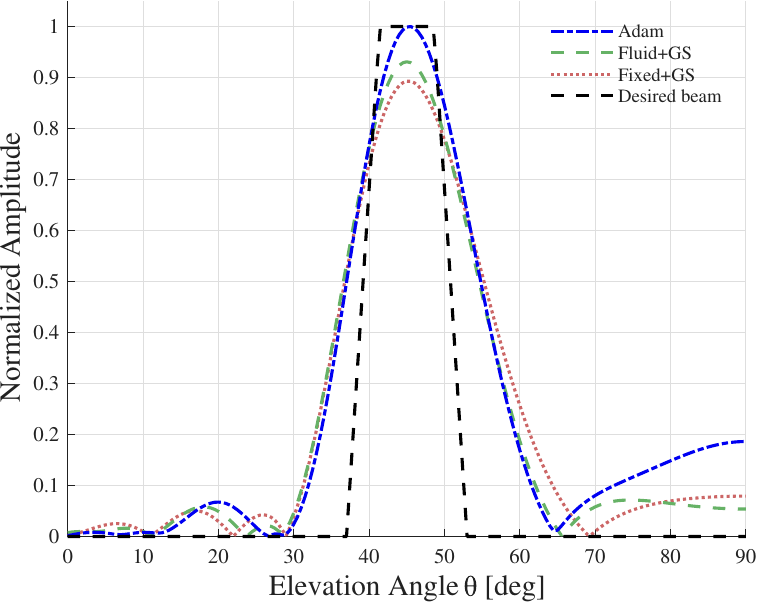}
	\caption{Continuous-stage normalized beam-amplitude cross section at
		\(\varphi=30^\circ\).}
	\label{fig:nr-cont-cross-theta}
\end{figure}

Figs.~\ref{fig:nr-cont-cross-phi} and~\ref{fig:nr-cont-cross-theta}
show the same comparison in two cross sections. Adam produces the
highest main peaks while maintaining the two-beam balance in the
\(\theta=45^\circ\) section. In the \(\varphi=30^\circ\) section,
it narrows the nearby shoulders and reduces the response outside the
mainlobe relative to Fluid+GS. The port and weight refinement thus
increases mainlobe concentration and improves the peak sidelobe level.

\section{Conclusion}
This paper extended the OMP-and-Fourier arbitrary-direction beamforming
framework for planar FAAs from discrete port selection
to continuous port-and-weight refinement. The beam synthesis problem was
formulated in the far-field spatial domain, where the Fourier structure
connects aperture-domain beamforming weights and beam-domain samples. A standard
OMP procedure was used to select active ports from a dense candidate
grid, and least squares on the selected support provided both the
discrete-stage beam synthesis and the initial complex beamforming weights for the
continuous stage. An FFT-based Gerchberg--Saxton phase retrieval step was further
introduced to construct a complex target beam from an amplitude-only
desired beam pattern, making the target more compatible with the finite
aperture.

Starting from the OMP placement and least-squares weights, projected
Adam jointly refined the continuous port coordinates and complex
weights through a loss combining mainlobe concentration and multi-beam
balance. Aperture projection and minimum-spacing repair maintained
feasible port positions, and the highest-concentration feasible
checkpoint provided the final design. In the evaluated two-beam setup,
the complete method increased the mainlobe energy fraction from
35.73\% to 48.92\% and reduced the PSLL from \(-2.30\)~dB to
\(-3.83\)~dB relative to the fixed array with the same aperture and
active-port count.
\balance

\end{document}